\documentstyle[12pt]{article}

\begin{document}

\begin{center}
 {\Large \bf  Remarks on semileptonic $B$ and $D$ decays into orbitally 
excited mesons}\\
\vspace{1.0cm}

{\large   H. B. Mayorga, A. Moreno Brice\~no and J. H. Mu\~noz}

 {\it Departamento de F\'{\i}sica, Universidad del Tolima,}\\
{\it A. A. 546, Ibagu\'e, Colombia}

\end{center}

\begin{abstract}
 We have obtained the differential decay rate and  calculated the 
branching ratios  of the exclusive semileptonic decays $B(D) \rightarrow 
Xl\nu$, where  $X$ is a $p$-wave meson, using the nonrelativistic ISGW 
quark model.    Our results are compared with the predictions of the ISGW2 
model. We have computed some branching ratios   that were not reported or 
were reported with $0.00$  in this model. For example, we find that  
$Br(B_c^- \rightarrow \overline{B_{s2}^{*0}}l^-\overline{\nu}) = 4.03 
\times 10^{-5}$, $Br(B_c^- \rightarrow \overline{B_2^{*0}}l^- 
\overline{\nu}) =3.65 \times 10^{-6}$ and $Br(D_s^+ \rightarrow f_2l^+\nu) 
= 2.7 \times 10^{-5}$,   which seems to be at the reach of forthcoming 
experiments.  Furthermore,  we have  classified  the  $B_{u,d,s} 
\rightarrow Tl\nu$  decays in two groups and  compared  the  semileptonic 
and nonleptonic decays  including a tensor meson in the final state.

\end{abstract}

\noindent
PACS number(s): 13.20.He, 13.20.Fc, 12.39.Jh

\vspace{2.5 cm}

\newpage

\noindent
1. {\bf Introduction}\\

We have calculated the differential decay rates and the numerical values 
of the branching ratios of the semileptonic  decays $B(D) \rightarrow 
Xl\nu$, where $X$ is an orbitally excited meson  $^3P_2$ (i.e.,   a tensor 
meson $T$) or  $^1P_1$, in the nonrelativistic quark model of 
Isgur-Scora-Grinstein-Wise (ISGW model) \cite{isgw}.  Specifically, we 
have considered exclusive semileptonic decays of the mesons $B_q$ ($q = 
u,d,s,c$) and  $D_{q^{'}}$ ($q^{'} = u, d, s$). Let us mention that the 
ISGW model \cite{isgw} shows the inclusive spectrum of the semileptonic 
$B$ and $D$  decays. So, in this work we have obtained the branching 
ratios to each exclusive channel  and compared them with the predictions 
of the ISGW2 model \cite{isgw2}.  We have computed some branching ratios   
that do not appear or appear  with $0.00$  in this model \cite{isgw2}.    
From our  results we can classify the $B_{u, d, s} \rightarrow Tl\nu$ 
decays in two groups and establish a connection between $B \rightarrow 
Tl\nu$ and $B \rightarrow TV$ ($V$ means a vector meson), which agrees 
with the Bjorken$^{'}$s relation if we ignore the nonfactorizable 
contributions.\\

Semileptonic $B$ decays have been studied widely. The current experimental 
data show that  semileptonic $B$ decays to ground  charmed mesons ($D$ and 
$D^*$) is approximately the $60\%$ of the inclusive semileptonic decay 
rate \cite{huang}-\cite{leibovich} and that the remainder $40\%$ should go 
to  excited charmed mesons and nonresonant final states 
\cite{huang}-\cite{drell}. Besides,  the semileptonic $B$ decays into 
$p$-wave charm mesons are the main  background in the measurement of the 
$Br(B \rightarrow D^*l\nu)$ \cite{babar}-\cite{massimo}. For these 
reasons, the production of orbitally excited charmed mesons (or $p$-wave 
charmed states) in semileptonic $B$ decays  is an interesting area of 
research and has been studied  in the literature in the frame of Heavy 
Quark Effective Theory (HQET) \cite{huang}, \cite{leibovich}, 
\cite{iw}-\cite{matsuda} and quark models \cite {isgw}, \cite{isgw2}, 
\cite{ebert}-\cite{morenas}. \\

There are four orbitally excited charmed states with $l=1$, which can be 
classified in two doublets: $(D_0, D_1^*)$ and $(D_1, D_2^*)$ with 
$j=1/2$, $J^P=(0^+, 1^+)$ and  $j=3/2$, $J^P=(1^+, 2^+)$, respectively.    
In this work, we have studied,  additional  to the semileptonic $B$ decays 
with  charmed mesons $B \rightarrow D_1(2420)l\nu$ and $B \rightarrow 
D_2^*(2460)l \nu$, all the exclusive channels $B_{u,d,s,c} \rightarrow 
Tl\nu$, where $T$ is a  $^3P_2$ orbitally excited tensor meson.  Let us 
point out that the ISGW model works well for the $B_c$ decays  since it is 
conformed by two heavy quarks.  The Heavy Quark Effective Theory is not an 
appropiate scenario when the two constituent quarks inside the meson are 
heavy  \cite{bc}. \\

Furthermore, we have computed the branching ratios of the semileptonic 
decays $D \rightarrow Tl\nu$. In this case, there is not enough  
theoretical and experimental information about them. The most of the 
branching ratios of these decays were reported with $0.00$ in the Ref. 
\cite{isgw2}. We think that forthcoming experiments at CLEO-c Project  
will allow to test some of these predictions in the nonrelativistic quark 
models \cite{isgw, isgw2}.  \\

\noindent
2. {\bf Differential decay rate of semileptonic decays with orbitally 
excited mesons in the final state}\\

Integrating the expression for $d^2\Gamma(P \rightarrow Xl\nu)/dxdy$, 
given in the ISGW model \cite{isgw}, we have obtained  the differential 
decay rate of any semileptonic decay of a pseudoscalar meson $P$,

\begin{equation}\label{semi}
\frac{d\Gamma(P \rightarrow Xl\nu)}{dt}= \frac{\left| V_{qq^{'}} \right|^2 
G_F^2}{32\pi^3m_P^2}\left| \vec{P}_X  \right| \left( \frac{4}{3}m_P^2 
\beta_{++} \left| \vec{P}_X \right|^2  + \alpha t \right),\\
\end{equation}

where $t = (p_P - p_X)^2$, $\vec{P}_X$ is the three-momentum of $X$ in the 
$P$ rest frame,  $G_F$ is the Fermi constant, $V_{qq^{'}}$ is the 
Cabibbo-Kobayashi-Maskawa (CKM) factor  and    $\beta_{++}$ and $\alpha$ 
are functions of the form factors \cite{isgw}. In the above equation $m_l 
\approx 0$ i.e., $l = e, \mu$.\\

Using the corresponding expressions for the functions $\beta_{++}$ and 
$\alpha$  given in the appendix of the Ref. \cite{isgw} when $X$ is an  
orbitally excited meson $^1P_1$, we obtain explicitly, from  the Eq. 
(\ref{semi}),  the following  differential decay rate of the semileptonic  
decay $B \rightarrow (^1P_1) l \nu$:

\begin{eqnarray}\label{orbitally1}
\frac{d\Gamma(B \rightarrow Xl^-\overline{\nu})}{dt} &=& \frac{\left| 
V_{qb} \right|^2 G_F^2}{96 \pi^3 m^2_{X}} \left\{  4 m_B^2 S_+^2 \left| 
\vec{P}_{X} \right|^5  + \left[ r^2 + 8m^2_{X}t v^2 + 2(m_B^2 - m_{X}^2 - 
t)rS_+ \right] \left| \vec{P}_{X} \right|^3   \right. \nonumber \\
&& \ \ \ \ \left.   + \frac{3m^2_{X}}{m^2_B}tr^2 \left|\vec{P}_{X}  
\right|  \right\},
\end{eqnarray}

where $S_+$, $r$ and $v$ are form factors given in the Ref. \cite{isgw} 
and $X$ is a $p-wave$ meson $^1P_1$.\\

On the other hand, we have taken  from the appendix of the Ref.  
\cite{isgw}    the corresponding expressions of the functions $\beta_{++}$ 
and $\alpha$   assuming that  $X$  is an orbitally excited $^3P_2$ tensor 
meson $T$.  In this case, we have obtained explicitly the following 
expression for the semileptonic decay rate of $B \rightarrow Tl\nu$

\begin{equation}\label{orbitally2}
\frac{d\Gamma(B \rightarrow Tl\nu)}{dt}=\frac{\left|V_{qb} 
\right|^2G_F^2}{288\pi^3 m^4_T}\left\{ \alpha(t) \left| \vec{P}_T  
\right|^7       + \beta(t) \left| \vec{P}_T  \right|^5 + \gamma(t) \left| 
\vec{P}_T  \right|^3  \right\},\\
\end{equation}

where $\alpha(t)$, $\beta(t)$ and $\gamma(t)$ are quadratic functions of 
the form factors  $b_+$, $k$ and $h$,  given by

\begin{eqnarray}
\alpha(t) &=& 8 m_B^4 b_+^2 ,\nonumber \\
\beta(t) &=& 2m_B^2 \left[6m_T^2th^2 + k^2 + 2(m^2_B - m_T^2 - t)kb_+     
\right] ,\nonumber \\
\gamma(t) &=& 5 m_T^2 t k^2.
\end{eqnarray}

From Eqs. (\ref{orbitally1}) and (\ref{orbitally2}) we can see that the 
decay width of the semileptonic decay $B(D) \rightarrow Xl\nu$, where $X$ 
is an orbitally excited meson $^1P_1$ or $^3P_2$ ( a tensor meson $T$) has 
three contributions. In one case, for $X=$ $^1P_1$, these contributions 
are proportional to  $| \vec{P}_X |^5$, $| \vec{P}_X  |^3$ and $| 
\vec{P}_X |$. On the other hand, when $X$ is a tensor meson $T$, they are 
proportional to $| \vec{P}_T |^7$,  $| \vec{P}_T |^5$ and  $| \vec{P}_T  
|^3$ (we call these contributions $\Gamma_{(7)}$, $\Gamma_{(5)}$ and 
$\Gamma_{(3)}$, respectively). It means
that the particles in the final state are coupled to waves $l= 2, 1, 0$ or 
$l=3, 2, 1$ when the orbitally excited meson is $^1P_1$ or $^3P_2$, 
respectively\footnote{We display in the table 2 the contributions 
$\Gamma_{(7)}$, $\Gamma_{(5)}$ and $\Gamma_{(3)}$ to the decay width of 
each exclusive channel $B \rightarrow Tl\nu$.}. \\

Now, we are going to  compare at tree level the semileptonic decay  
$B(b\overline{q}) \rightarrow  T(q^{'}\overline{q})l\nu$ with the {\it 
type-I} nonleptonic decay  $B(b\overline{q}) \rightarrow 
T(q^{'}\overline{q})V(q_i\overline{q_j})$, which  are produced by the 
current matrix element $<T|J_{\mu}|B>$ (i.e.  the tensor meson $T$ is 
produced from the transition $B \rightarrow T$).  If we suppose that the 
only factorizable contribution to $B \rightarrow TV$  comes from 
$<T|J_{\mu}|B><V|J^{\mu}|0>$ and neglect the nonfactorizable 
contributions, we can 
establish a ratio between the Eq. (\ref{orbitally2}) and the Eq. (11) of 
the Ref. \cite{herman}. It is given by

\begin{equation}\label{comparison}
{\cal R}\equiv \frac{\Gamma(B \rightarrow TV)}
{\left. \frac{d\Gamma(B \rightarrow Tl\nu)}{dt}\right|_{t=m_V^2}
}=6 \pi^2 \left|V_{ij} \right|^2a_1^2F_V^2,\\
\end{equation}

where  $F_V = m_Vf_V$ is the decay constant ($f_V$ is adimensional in this 
case), $V_{ij}$ is the appropiate CKM factor (depending on the flavor 
quantum numbers of the meson $V$) and $a_1$ is the QCD coefficient. \\

In the literature, neglecting nonfactorizable contributions to the decay 
amplitude, is well known  that an identical expression is obtained for  
the ratio  ${\cal R}$  when we compare the decays $B \rightarrow Xl\nu$ 
and $B \rightarrow XV$,  $X$ being a pseudoscalar ($P$) or a vector ($V$) 
meson and arising from the transition $B \rightarrow X$. Therefore, we 
would like to point out that in this work we show explicitly that the  
ratio ${\cal R}=6 \pi^2 \left|V_{ij} \right|^2a_1^2F_V^2$ between $B 
\rightarrow Xl\nu$ and $B \rightarrow XV$  is always the same irrespective 
of the meson $X$ is a pseudoscalar, a vector or a tensor meson. It means 
that the production of the lepton pair in the semileptonic decay is 
kinematically equivalent to the production of a pseudoscalar or a vector 
\cite{babar, neubert-stech} or a tensor meson in the nonleptonic decay. It 
is important to note that the expression for  ${\cal R}$, which is 
independent of the model, also agrees with the Bjorken$^{'}$s relation 
\cite{babar, neubert-stech, bjorken} and it provides  a clear test of the 
factorization hypothesis and may be employed to determine unknown decay 
constants \cite{babar, neubert-stech}. \\

\noindent
3. {\bf Numerical values}\\

In this section we calculate the branching ratios of the exclusive 
semileptonic decays $B_q (D_{q^{'}}) \rightarrow Tl\nu$, where $T$ is an 
orbitally excited  meson $^3P_2$, $q=$ $u,d,s,c$ and $q^{'}=u,d,s$, and 
$B^-(\overline{B_s^0}) \rightarrow D_1^0(D_{s1}^+) l^- \overline{\nu}$ 
using the ISGW model \cite{isgw}.  In order to provide numerical values of 
the branching ratios of $B \rightarrow$ $^1P_1 l\nu$ and $B(D) \rightarrow 
T l\nu$ we use the expressions for the differential decay rates given in 
Eqs. (\ref{orbitally1}) and (\ref{orbitally2}), respectively,   and  the 
following values of the  CKM elements \cite{pdg}: $|V_{cb}|=0.0402$, 
$|V_{ub}|= 3.3 \times 10^{-3}$, $|V_{cs}|=0.97$, $|V_{cd}|=0.224$. The 
values for the lifetime of $B_{u, d, s, c}$ and $D_{u, d, s}$ and all the 
masses required are taken form \cite{pdg}. We have used an octet-singlet 
mixing angle $\theta_{T}=28^{\circ}$ when we computed the decays that  
involve an isoscalar tensor meson $f_2$ or $f_2^{'}$.  \\

In the table 1 we present our results (see the second column) for the 
branching ratios of $B \rightarrow Tl\nu$ and $B^-(\overline{B_s^0}) 
\rightarrow D_1^0(D_{s1}^+) l^- \overline{\nu}$ ($D_1^0$ and $D_{s1}$ are 
$^1P_1$ mesons). In the third (fourth) column we display the values 
obtained from the ISGW2 model \cite{isgw2} (other references in the frame 
of quark models or HQET). We can see that, in general, the values obtained 
in the ISGW model \cite{isgw} are smaller than the values obtained in the 
ISGW2 model \cite{isgw2}.  A similar conclusion was obtained recently by 
\cite{kim} working with nonleptonic two-body $B$ decays. The experimental 
values of CLEO \cite{pdg, cleo} $Br(B^-  \rightarrow D_2^{*0}l^- 
\overline{\nu}) < 8 \times 10^{-3}$ and $Br(B^- \rightarrow D_1^0 l^- 
\overline{\nu}) = (5.6 \pm 1.6) \times 10^{-3}$ agree with all the 
predictions obtained from the quark models  \cite{isgw, isgw2, ebert, 
morenas}  and the HQET \cite{huang, wang, dai}.\\

Let us mention that we obtain ${\cal R}= [Br(B^- \rightarrow 
D_2^{*0}l^-\overline{\nu})/Br(B^- \rightarrow D_1^0l^- \overline{\nu})]= 
0.3539$ which is consistent with the experimental limit ${\cal R} < 1.42$ 
\cite{pdg} and the report of the Ref. \cite{battaglia}.   On the other 
hand  the sum of all the branching ratios of $B_{u,d} \rightarrow Xl\nu$ 
(where $X$ is an orbitally excited meson),  calculated in this work (see 
the second column in the table 1),  is almost   0.745\% (i. e. $1.49\%$ 
including the conjugate channels). \\

In the table 2 we show the three contributions $\Gamma_{(7)}$, 
$\Gamma_{(5)}$ and $\Gamma_{(3)}$,  given by the Eq. (\ref{orbitally2}), 
to the decay width of $B \rightarrow Tl\nu$. We note that the $B_{u, d, s} 
\rightarrow Tl\nu$ decays can be classified in two groups. In one of them, 
all the contributions are positive and  the contribution $\Gamma_{(3)}$, 
which is proportional to  $| \vec{P}_T |^3 $  is the largest.  The decays 
which come from the  $b \rightarrow u$ transition are in the other group. 
In this case, the  contribution $\Gamma_{(7)}$, which is proportional to 
$| \vec{P}_T |^7 $,  is the largest  and the contribution $\Gamma_{(5)}$, 
which is proportional to   $| \vec{P}_T |^5 $,  is negative. For $B_c 
\rightarrow Tl\nu$  these contributions, in all the cases,  are positive 
and $\Gamma_{(3)}$ is the largest. \\

Finally, in the table 3 we show the numerical values of the branching 
ratios for the decays $D_{q^{'}} \rightarrow Tl\nu$ ($q^{'}= u, d, s$).
In this case there is not enough theoretical predictions and experimental 
data. So, we have only compared our results  with the ISGW2 model 
\cite{isgw2}. The most of the branching ratios were reported with $0.00$ 
in this Ref. \cite{isgw2}. The $Br(D^0  \rightarrow a_2^- l^+ \nu)$ and 
$Br(D_s^+  \rightarrow K_2^{*0} l^+ \nu$) are enhanced by about an order 
of magnitude in the ISGW2 model. Our prediction for $Br(D^+  \rightarrow 
\overline{K_2^{*0}}l^+ \nu)$ is below three orders of magnitude of the 
experimental limit  ($< 8 \times 10^{-3}$) \cite{pdg}. Let us mention  
that we do not display the contributions $\Gamma_{(7)}$, $\Gamma_{(5)}$ 
and $\Gamma_{(3)}$, given  by the Eq. (\ref{orbitally2}), to $\Gamma(D 
\rightarrow Tl\nu)$ because their behaviour is similar to the case of $B_c 
\rightarrow Tl\nu$ decays. \\

 \noindent
4. {\bf Conclusion}\\

We have computed the branching ratios of the semileptonic $B$ and $D$ 
decays involving $^1P_1$ and $^3P_2$ mesons in the final state (they are  
$p$-wave orbitally excited mesons) using the nonrelativistic quark ISGW 
model \cite{isgw}. We have calculated some branching ratios that were not 
predicted or were reported with $0.00$ or  in the Ref. \cite{isgw2}. For 
example,  we have found that $Br(B_c^- \rightarrow 
\overline{B_{s2}^{*0}}l^-\overline{\nu}) = 4.03 \times 10^{-5}$, $Br(B_c^- 
\rightarrow \overline{B_2^{*0}}l^- \overline{\nu}) =3.65 \times 10^{-6}$ 
and $Br(D_s^+ \rightarrow f_2l^+\nu) = 2.7 \times 10^{-5}$, which seems to 
be within  the reach of forthcoming experiments. In general, the branching 
ratios predicted by the ISGW model are smaller than the predictions of the 
ISGW2 model. At this time, the predictions for $B^-  \rightarrow 
D_2^{*0}l^- \overline{\nu}$ and $B^- \rightarrow D_1^0 l^- \overline{\nu}$ 
 from  the quark models and HQET agree with the experimental data.  \\

The decays $B_{u, d, s} \rightarrow Tl\nu$ can be classified in two 
groups, depending on  the contributions $\Gamma_{7}$,  $\Gamma_{5}$ and   
$\Gamma_{3}$, given by the Eq. (\ref{orbitally2}), to the   decay width. 
We also establish a relation between the semileptonic decay $B \rightarrow 
Tl\nu$ and the nonleptonic decay $B \rightarrow TV$, neglecting  the   
nonfactorizable and the annihilation contributions.   \\

\vspace{0.7cm}

{\it Acknowledgements}  We wish to express our gratitude   to  G. L\'opez 
Castro for reading the manuscript and his valuable suggestions and the 
{\it Comit\'e Central de Investigaciones} (CCI) of the University of 
Tolima  for financial support.

\vspace{1cm}

\begin{center}
\begin{tabular}{||c|c|c|c||}
\hline
 Process &  This work & ISGW2 \cite{isgw2}&  Other Refs.\\
\hline\hline
$B^-  \rightarrow D_2^{*0}l^- \overline{\nu}$ &  $1.56 \times 10^{-3}$ 
&$2.4 \times 10^{-3}$ & $5.2 (5.9) \times 10^{-3}$  $m_Q \rightarrow 
\infty $ $ (1/m_Q)$ \cite{huang, dai}\\ \cline{4-4}
 & & & $(7-4.7-6.5-7.7) \times 10^{-3}$ \cite{morenas} \\ \cline{4-4}
& & & $(7.9 \pm 2.3) \times 10^{-3}$ \cite{wang}\\ \cline{4-4}
& & & $5.2 \times 10^{-3}$  \cite{ebert} \\ \cline{4-4}
\hline
$B^-  \rightarrow a_2^0l^- \overline{\nu}$ &  $6.31 \times 10^{-6}$ & 
$2.88 \times 10^{-5}$ &
\\
\hline
$B^-  \rightarrow f_2l^- \overline{\nu}$ &  $5.32 \times 10^{-6}$
& $3.24 \times 10^{-5}$ & \\
\hline
$ B^-  \rightarrow f_2^{'}l^- \overline{\nu} $ &  $9.56 \times 10^{-7}$& &
\\
\hline
$\overline{B^0}  \rightarrow D_2^{*+}l^- \overline{\nu}$ & $1.46 \times 
10^{-3}$ & $2.25 \times 10^{-3}$  &
\\
\hline
$\overline{B^0}  \rightarrow a_2^{+}l^- \overline{\nu}$ &  $1.18 \times 
10^{-5}$ & $5.56 \times 10^{-5}$ &
\\
\hline\hline
$\overline{B_s^0}  \rightarrow D_{s2}^{*+}l^- \overline{\nu}$ & $1.76 
\times 10^{-3}$ & $3.13 \times 10^{-3}$ & $5.9 \times 10^{-3}$  
\cite{ebert}
\\
\hline
$\overline{B_s^0}  \rightarrow K_{2}^{*+}l^- \overline{\nu}$ &  $6.61 
\times 10^{-6}$ & $4.55 \times 10^{-5}$ &
\\
\hline\hline
$B_c^-  \rightarrow \chi_{c2}l^- \overline{\nu}$ &  $6.07 \times 10^{-4}$ 
& $8.17 \times 10^{-4}$ & $1.407 (1.909) \times 10^{-3}$  \cite{chang}
\\
\hline
$B_c^-  \rightarrow \overline{D_2^{*0}} l^- \overline{\nu}$ &  $7.57 
\times 10^{-7}$ & $3 \times 10^{-6}$ &
\\
\hline
$B_c^-  \rightarrow \overline{B_{s2}^{*0}} l^- \overline{\nu}$ &  $4.03 
\times 10^{-5}$ & $0.00$ & 
\\
\hline
$B_c^-  \rightarrow \overline{B_2^{*0}} l^- \overline{\nu}$ & $3.65 \times 
10^{-6}$ & &
\\
\hline\hline\hline\hline
$B^- \rightarrow D_1^0 l^- \overline{\nu}$ & $4.407 \times 10^{-3}$ & $4.8 
\times 10^{-3}$ & $3.4 (13)\times 10^{-3}$ $m_Q \rightarrow \infty$   
$(1/m_Q)$  \cite{huang, dai}\\ \cline{4-4}
& & &  $(4.5-2.9-4.2-4.9) \times 10^{-3}$ \cite{morenas}     \\ \cline{4-4}
& & & $(4.5 \pm 1.3) \times 10^{-3}$  \cite{wang}        \\ \cline{4-4}
& & & $3.3 \times 10^{-3}$  \cite{ebert}\\
\hline
$\overline{B_s^0} \rightarrow D^+_{s1}l^- \overline{\nu}$ & $1.95 \times 
10^{-3}$ & $5.3 \times 10^{-3}$ & $3.9 \times 10^{-3}$  \cite{ebert}\\
\hline\hline
\end{tabular}
\end{center}
\begin{center}
Table 1. Branching ratios of the semileptonic $B$ decays with orbitally 
excited mesons $^3P_2$ and $^1P_1$ (see the last two processes).
\end{center}

\vspace{1cm}

\begin{center}
\begin{tabular}{||c|c|c|c||}
\hline
Process & $ \Gamma_{(7)} $ & $\Gamma_{(5)}$ & $\Gamma_{(3)}$  \\
\hline\hline
$B^-  \rightarrow D_2^{*0}l^- \overline{\nu}$ & $1.2627$ & $1.4648$  & 
$3.4873$ 
\\
\hline
$B^-  \rightarrow a_2^0l^- \overline{\nu}$ & $4.8092 \times 10^{-2}$  & $ 
-3.4228 \times 10^{-2}$ & $1.1283 \times 10^{-2}$
\\
\hline
$B^-  \rightarrow f_2l^- \overline{\nu}$ & $4.1825 \times 10^{-2}$ & 
$-3.0097 \times 10^{-2}$ & $9.4626 \times 10^{-3}$
\\
\hline
$ B^-  \rightarrow f_2^{'}l^- \overline{\nu} $ & $6.3540 \times 10^{-3}$ & 
$-4.2931 \times 10^{-3}$ & $1.747 \times 10^{-3}$
\\
\hline
$\overline{B^0}  \rightarrow D_2^{*+}l^- \overline{\nu}$ & $1.2638 $& 
$1.4646$ & $3.4893$
\\
\hline
$\overline{B^0}  \rightarrow a_2^{+}l^- \overline{\nu}$ & $9.6234 \times 
10^{-2}$ & $-6.8486 \times 10^{-2}$ & $2.2575 \times 10^{-2}$
\\
\hline\hline
$\overline{B_s^0}  \rightarrow D_{s2}^{*+}l^- \overline{\nu}$ & $1.1022  $ 
& $2.3838$ & $4.3102$
\\
\hline
$\overline{B_s^0}  \rightarrow K_{2}^{*+}l^- \overline{\nu}$ & $2.3412 
\times 10^{-2} $ & $-1.2501 \times 10^{-2}$ & $1.8244 \times 10^{-2}$
\\
\hline\hline
$B_c^-  \rightarrow \chi_{c2}l^- \overline{\nu}$ & $4.3443 \times 10^{-1}$ 
& $3.0763$ & $5.1773$
\\
\hline
$B_c^-  \rightarrow \overline{D_2^{*0}} l^- \overline{\nu}$ & $6.7408 
\times 10^{-4} $ & $2.1651 \times 10^{-3}$ & $8.0017 \times 10^{-3}$ 
\\
\hline
$B_c^-  \rightarrow \overline{B_{s2}^{*0}} l^- \overline{\nu}$ & $2.3402 
\times 10^{-3} $ & $2.5878 \times 10^{-1} $ & $3.1576 \times 10^{-1} $
\\
\hline
$B_c^-  \rightarrow \overline{B_2^{*0}} l^- \overline{\nu}$ & $3.0461 
\times 10^{-4} $ & $2.2567 \times 10^{-2} $ & $2.9442 \times 10^{-2} $
\\
\hline\hline
\end{tabular}
\end{center}
\begin{center}
Table 2. Contributions to $\Gamma(B \rightarrow T l \nu)$ proportional to  
 $| \vec{P}_T |^7$, $| \vec{P}_T |^5$ and $| \vec{P}_T |^3$. All the 
values must be multiplied by $10^{-16}$.
\end{center}

\vspace{1cm}

\begin{center}
\begin{tabular}{||c|c|c||}
\hline
 Process & This work & ISGW2 \cite{isgw2}\\
\hline\hline
$D^+  \rightarrow \overline{K_2^{*0}}l^+ \nu$ &  $7 \times 10^{-6}$  
&$0.00$\\
\hline
$D^+  \rightarrow a_2^0l^+ \nu$ & $5.8 \times 10^{-7}$ & $0.00$\\
\hline
$D^+  \rightarrow f_2 l^+ \nu$ &  $7.65 \times 10^{-7} $ & $0.00$\\
\hline
$D^+  \rightarrow f_2^{'} l^+ \nu$ &  $ 4.72 \times 10^{-9}$ &
\\
\hline
$ D^0 \rightarrow K_2^{*-} l^+ \nu $ & $2.86 \times 10^{-6} $ &$0.00$ \\
\hline
$D^0  \rightarrow a_2^- l^+ \nu$ &  $ 4.32 \times 10^{-7}$ & $2.07 \times 
10^{-6}$ \\
\hline\hline\hline
$D_s^+  \rightarrow f_2 l^+ \nu$ &  $2.7 \times 10^{-5}$ & \\
\hline
$D_s^+  \rightarrow f_2^{'} l^+ \nu$ &  $4.03 \times 10^{-6}$  & $0.00$\\
\hline
$D_s^+  \rightarrow K_2^{*0} l^+ \nu$ &  $6.16 \times 10^{-7} $ & $2.48 
\times 10^{-6}$\\
\hline\hline
\end{tabular}
\end{center}
\begin{center}
Table 3. Branching ratios of the decays $D \rightarrow Tl\nu$.
\end{center}

\end{document}